# AI and Social Theory

Jakob Mökander[1], Ralph Schroeder[1]

[1] Oxford Internet Institute, University of Oxford, 1 St Giles, Oxford OX1 3JS, United Kingdom

Email for correspondence < jakob.mokander@oii.ox.ac.uk >

## Abstract

In this paper, we sketch a programme for AI-driven social theory. We begin by defining what we mean by artificial intelligence (AI) in this context. We then lay out our model for how AI-based models can draw on the growing availability of digital data to help test the validity of different social theories based on their predictive power. In doing so, we use the work of Randall Collins and his state breakdown model to exemplify that, already today, AI-based models can help synthesize knowledge from a variety of sources, reason about the world, and apply what is known across a wide range of problems in a systematic way. However, we also find that AI-driven social theory remains subject to a range of practical, technical, and epistemological limitations. Most critically, existing AI-systems lack three essential capabilities needed to advance social theory in ways that are cumulative, holistic, open-ended, and purposeful. These are (1) semanticization, i.e., the ability to develop and operationalize verbal concepts to represent machine-manipulable knowledge; (2) transferability, i.e., the ability to transfer what has been learned in one context to another; and (3) generativity, i.e., the ability to independently create and improve on concepts and models. We argue that if the gaps identified here are addressed by further research, there is no reason why, in the future, the most advanced programme in social theory should not be led by AI-driven cumulative advances.

## Key words

## 1. Introduction

In this paper, we sketch a programme for AI-driven social theory. First, we define what we mean by artificial intelligence (AI) in this context. We go on to outline a model for how AI-systems can leverage the growing availability of data to test hypotheses and generate new theories. Next, we give an example where computer-aided theory has already been put forward and tested; namely, Collins' state breakdown model. We then discuss the strengths and gaps of current AI techniques in relation to the task of advancing social theory. Finally, we point to some obstacles that lie ahead, as well as to reasons why AI-driven social theory is nevertheless likely to progress. We conclude with some implications for social science.

Before we begin, a quite abstract question should be addressed: what is social theory good, or useful, for? One purpose of theory is to guide knowledge cumulation (Schroeder 2019). It does so by synthesizing what is known in terms of generalizations in the social sciences. However, a review of the field a quarter of a century ago identified one of the problems with this view, which can be quoted in full: 'The general public, to the extent that it has any opinion of social theory at all, probably considers it to be mere ideology. So long as theories are mere meditations punctuated with dubious metaphors, there is little defense against this accusation. ASI [Artificial Social Intelligence] and mathematical formalism are compatible methods for stating a theory precisely, connecting its concepts in rigorous intellectual structures, and identifying both hidden assumptions and unexpected consequences' (Bainbridge et al. 1994: 431). Another aim of linking social theory and AI could be to focus knowledge cumulation on particularly acute social problems and how to tackle them. We will return to the question of how social theory can be applied to social problems on several occasions. But whether social theory is primarily problem-driven or driven by cumulation, it always focuses on gaps in knowledge and moving knowledge beyond what is known, which, as we shall see, overlaps with the aims of AI research.

Despite the shared commitment to cumulate knowledge, however, there are also differences between AI research and social science. One is that, in comparison with the natural world, social science or social theory comes up against allegedly greater complexity in dealing with the social world. This complexity is sometimes linked to the reflexivity of social science, whereby the very observation of society affects and is affected by knowledge. Another facet of this complexity is the inherent unpredictability of social phenomena. Yet both of these ideas are misguided: social theory can be predictive, and the social world is separate from observers, as is the natural world. It is not the complexities or reflexive ties between knowledge and social development that limit social theory. There are, however, other limits to our ability to model the social world, such as how well various kinds of models fit the social world (see Page 2018) and how problems in social theory such as the macro- micro- linkage can be tackled (to be discussed below). Though obviously not the only way forward, a major opportunity to push the boundaries of the aforementioned limitations, as we shall see, is to incorporate previous social science knowledge in AI-based models and to harness AI for knowledge cumulation.

A number of oppositions are useful to keep in mind as we proceed. One is, again, between macro-micro- sociology. Further oppositions include top-down versus bottom-up approaches in AI (Mitchell 2019), inductive versus deductive reasoning, and data-driven versus theory-driven knowledge cumulation. Note the parallel oppositions in AI research and in the social sciences here, as when top-down versus bottom-up approaches to AI have an echo in micro- versus macro- social relations. Top-down AI hinges on the belief that intelligent behaviour can be programmed into systems from scratch using logic and symbolic representations. But similar approaches could also be generated bottom-up, by means of a plethora of (interacting) autonomous devices. For social science, the top-down direction of travel is also a society steered by formalized, universal knowledge – technocracy – while again, steering social processes could also emerge bottom-up, through the interactions of knowledge-steered social actors. A related opposition that applies to both AI research and the social sciences is whether there is too much or too little data: too much is not often considered in either domain, but we shall come back to this point.

One final opposition worth considering is that between computer science and social science. In this paper, computer science will only be touched on briefly, when we discuss how the technical capabilities of different types of AI-systems fit with the aim of contributing to advancing social theory. However, because the assumptions and values of computer scientists help shape the affordances of the AI-systems they develop, it is worth mentioning some disciplinary differences. Computer scientists are typically not versed in social theory and they often use natural science theories such as evolution or complexity in relation to social science phenomena. However, theories of social phenomena must fit with cumulative social science – and this fit is often limited for natural science theories. In a similar way, social scientists are typically not versed in advanced computational techniques such as machine learning (ML) or neural networks (Molina and Garip 2019). This lack of technical literacy can lead to unsubstantiated fears about the social challenges posed by AI on the one hand, as well as unrealistic expectations about what AI-systems can reasonably be expected to achieve on the other.

Let us now return to sketching the programme for AI-driven social theory. Knowledge cumulation is currently impeded because research within AI and social theory remain separate enterprises. This observation is not new: already in 1994, Bainbridge et al. argued that a strategic use of AI-systems can not only render existing social theories more rigorous, but also inspire new theories. Since then, much progress has been made, and, in more recent years, AI-based models have been used by social scientists to study individuals and groups (Hoey et al. 2018), to develop theories of human behaviour (Edelmann et al. 2020) as well as to evaluate results previously obtained by means of traditional statistical techniques (Di Franco and Santurro 2020). However, as Radford and Joseph (2020) point out, work in this area too often privileges AI-based models that perform well on narrowly defined technical parameters over models that are founded in a deeper understanding of the society under study. Hence, there is a fine balance between applying AI-based models to social data to advance our understanding of the social world, on the one hand, and of falling into a pseudoscience, where misappropriated algorithms are deployed to make baseless social scientific claims, on the other.

The solution we propose here is to spell out a coherent programme of research which consists of unified (seamless from micro- to macro-), consistent, and coherent advances in social theory by AI-systems, i.e., those that are both state-of-the-art and generalizable. We have chosen to present this programme in terms of a single thinker (Collins), but that thinker can be seen as a cipher; Collins is simply drawing together a number of strands in social theory. This is not new or unconventional: In the social sciences, theories are often associated with thinkers such as Marx, Weber, or Durkheim. Even theories in theoretical traditions such as social network analysis are associated with certain figures such as Harrison White (1992), though there also accounts focusing on more on theory development in social network analysis than on thinkers (Freeman 2004; Rule 1997: 120-43). But focusing on thinkers is misleading, as Collins and (also Weber) would be the first to point out: it is not thinkers but useful theories that drive knowledge advancements, even if individual figures provide a heuristic shorthand for those theories. Hence, there is no reason why, in the future, the most advanced programme in social theory should not be associated with the name of a software programme rather than the name of a social thinker. Against this background, we can begin by defining what we mean by AI in this context.

## 2. Operationalizing AI to advance social theory

There is no widely accepted definition of AI. Rather, practitioners across different disciplines define, interpret, and operationalize AI in quite different ways. However, it is possible to discern three main ways of understanding the term: AI can refer to either: a) an academic discipline, i.e., a branch of computer science; b) an agent that is characterized by high levels of autonomy, adaptability, and other specific competencies; or c) a set of technologies or tools and methods that collect and process data. John McCarthy, an early AI pioneer who is often credited with having coined the term, mainly championed the first of these conceptualizations. According to McCarthy, AI is the science and engineering of making intelligent machines (McCarthy 2007). The second way of conceptualizing AI can be illustrated by reference to Wang (1995), who focuses on the ability of an intelligent agent to 'adopt to its environment with insufficient knowledge and resources', or to Albus (1991), who stipulates that an intelligent agent is one that can 'act appropriately so as to increase the probability of achieving complex goals'. However, neither of these conceptualisations correspond to what we mean when referring to AI in this paper: we leave it to computer scientists to build intelligent machines. Instead, we are concerned with how automated computational techniques, fueled by the growing availability of big data (Brady 2019), can help us to better model and understand social change.

For our purposes, then, we define AI as a system (i.e., a set of technologies) that collects and processes data in order to refine a model so as to accurately represent the social world. Such AI-systems learn about their environment through automated data acquisition, interpret the collected data by means of computational techniques, and represent patterns either symbolically or visually, thus contributing to the cumulation of social scientific knowledge via better fitting models of social change. This approach is in line with the third way of conceptualizing AI above, which is also how policymakers tend to view AI. The US Department of Defense, for example, defines AI as a set of ML techniques that are designed to approximate cognitive tasks (US DoD 2018). The cognitive task, in our case, is to identify patterns in the social world and link these until they form an interconnected system on the basis of which social theory can advance. For such AI-based models to be feasible and effective, a number of technical affordances are required. According to Russell and Norvig (2015), the most important capabilities can be summarized as follows: first, the AI-system needs sensors to capture data; second, it needs knowledge representations to store the data; third, it needs automated reasoning to use the stored data to draw new conclusions; fourth, it needs the ability to learn, both to adapt to new evidence and to detect patterns; and fifth, it needs interfaces to communicate, either with other machines or with human operators.

Given these a priori requirements, it is evident that recent advances in computer science research will impact the ability of future AI-systems to advance social theory. While technologies like computer vision unlock new data sources, automated web crawling and big data analytics help make sense of data where they are already available (Mayer-Schönberger and Cukier 2013). Similarly, increased storage capacity and processing power both enable more sophisticated ways to represent knowledge and allow for heuristic programs to conduct more exhaustive searches (Mitchell 2019). Perhaps most importantly, advances in ML techniques – notably deep learning (Goodfellow et al. 2016) and reinforcement learning (Sutton and Barto 2018) – allow AI-systems to learn autonomously, i.e., unsupervised. The ability of deep neural networks to draw non-intuitive inferences from unstructured or structured data means that AI-systems can not only be used to test hypotheses but also generate new hypotheses. It should be noted here that deep learning and reinforcement learning are not mutually exclusive. In fact, deep neural networks may be trained using a reinforcement learning approach (Francois-Lavet et al. 2018). Reinforcement learning, in turn, is particularly well suited for the purpose of advancing social theory since it does not require a separate training set but can learn dynamically by adjusting its actions over multiple iterations so as to maximize a reward function (e.g., model fit with empirical evidence). Finally, advances in natural language processing and data visualization help AI-systems to represent their models to, and communicate with, human operators by providing new or enhanced interfaces.

AI-systems, as defined above, are inevitably part of larger sociotechnical systems. Thus, both the purpose for which an AI-system is applied and the quality of the model it generates are functions of many factors, including the data used and the sociotechnical environment surrounding its use. Consequently, even when the available technologies afford AI-systems to assist the process of knowledge creation, they do not automatically do so. Human researchers and operators have to facilitate the process by providing a beneficial learning environment and a sound protocol for advancing theory. Here, we offer four principles for good AI design (adopted from Mökander 2021). These stipulate that AI-based models should be: (1) cumulative, i.e., advance knowledge by improving upon the synthesis of previous knowledge. Importantly, this also implies that the model is abstracted to fit the evidence most comprehensively and parsimoniously, based only on scientific criteria; (2) holistic. This means that they seek and integrate knowledge across all domains and apply its criteria globally and uniformly; (3) open-ended, i.e., not exclusive in allowing for the inclusion of yet more, and improved, data sources and ways to incorporate them; and (4) purposeful. In the case of advancing social theory, purposeful means that AI-based models should help social scientists either better understand or develop new methodologies for researching the dynamics of social change.

Two final remarks help complete our framing of AI. First, in previous literature, AI is interchangeably referred to as 'machine intelligence' or 'synthetic intelligence' (Wang 2019). In this paper, however, we will exclusively use the term AI for the simple reason that it is easier to communicate with one rather than multiple names. Second, given the cognitive nature of theorizing, our suggestion that AI-based models can help advance social theory may spark an age-old question, namely: is it possible for computing machines to think? But for the programme we sketch here, the answer to this question is less important. If by thinking we mean the activity and experience peculiar to humans, then the answer is a resounding no. However, even human actions can be understood in terms of interlocking feedback loops in which signals between subsystems result in complicated but stable behavior (Wiener 1954). Thus, if the question refers to experiment and empirical observation, then surely it is possible for machines to display thinking-like behaviour. For our practical purposes, and to advance social theory, that is quite enough.

## 3. AI-driven Social Theory: what it is and how it works

So far, we have argued that the combination of AI-based models and new sources of digital data can help us develop, test, and advance social theory. But how could AI-based models help advance social theory in practice? And how could AI-systems help bridge theories about different social domains, as well as the micro- macro- divide? In this section, we attempt to answer these questions schematically.

Figure 1 (below) illustrates the most important aspects of AI-driven social theory: The distillation of cumulated available theory plus data sources are put together in a hypothesis by visualizing and/or formalizing causal relations into a theory. Explanations are then ranked (or eliminated) by reference to their fit with evidence or data – whence, if they are deemed sufficient, social theory moves onto new terrain. If the fit is deemed incomplete, then a new hypothesis is generated that either extends the breadth or depth of the existing theory or uses new evidence or data.

In either case, these new elements are semanticized and enter into a new iteration of hypothesis formulation, so that the AI-based model learns in an ongoing iterative process. Figure 1 thereby illustrates how an AI-based model can help find the best explanation for a specific social phenomenon or structure. However, to be truly generalisable, such explanations or theories must not only hold true for each known special case, but also be mutually compatible (although some theories accommodate each other by synchronized non-simultaneity). Thus, apart from learning specific social phenomena, the programme we outline here is also designed to ensure that explanations in one domain are compatible with, and transferable to, other domains or social processes.

This leads to figure 2 (below), which shows the connections between various domains and the operations performed on them at a system level. By aggregating AI-based models of specific social phenomena, the AI-system contains a global model of the social world. Since this model is maximized for fit, it must incorporate all subsystemic models and their degrees of fit. Howsoever many of these subsystemic levels there are (one level down from the global level there are conventionally three such systems – the economic, political, and cultural orders or spheres; see Schroeder 2013), these macro- level models, and, downwards, all subsystems on the meso- and micro- levels, must be aggregated for best fit into the overall model. In any event, the mode of operation displayed in figure 1, i.e., where explanations are ranked through an iterative learning process, is applicable to all the subsystems in figure 2. The key is that the AI-system is optimizing, i.e., maximizing for fit with evidence, on a systemic level, thereby imposing constraints on individual AI-based models – whereby existing social theory is updated and advanced. Due to their generativity, the AI-based models can also identify and suggest new concepts or structures through which the social world can be understood (more on this in section 5). But existing theory would only be expanded if the inclusion of new concepts helps the AI-system maximize overall fit.

*Figure 1. AI-based model of social phenomenon (X) (Sub-system)*

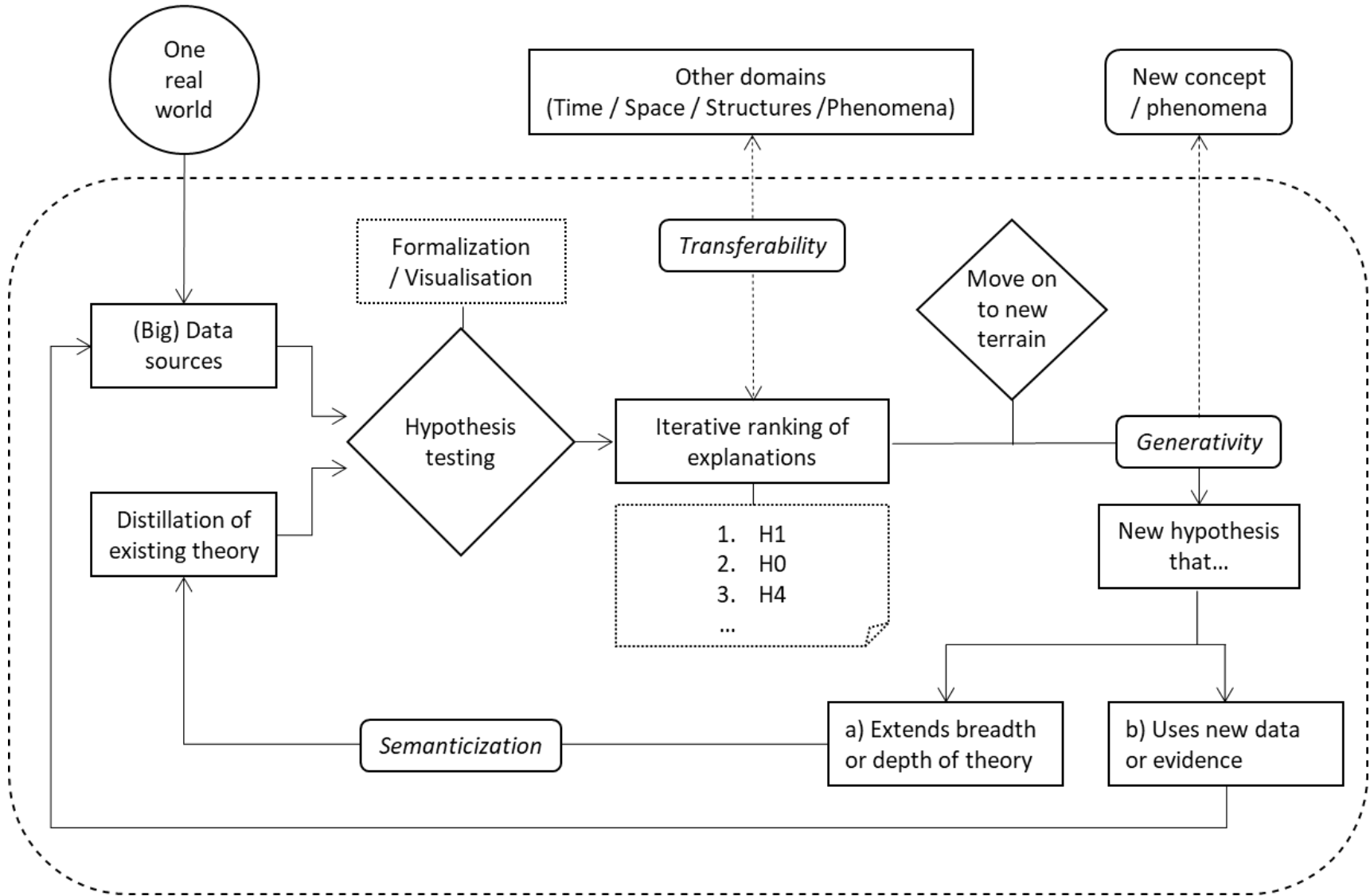

Figure 2. AI-System (Aggregating AI-based-models of social phenomena)

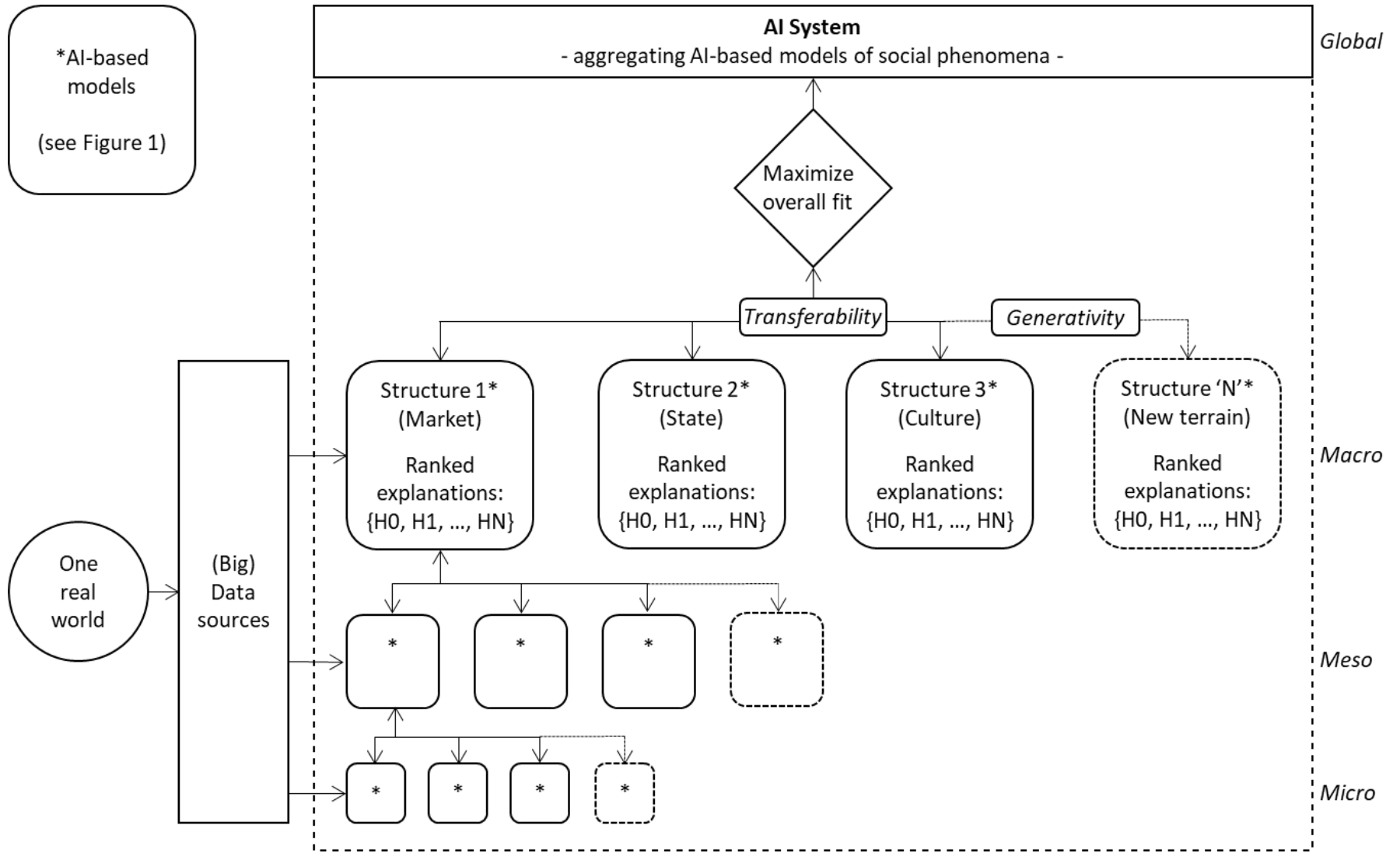

## 4. AI and Social Theory: the example of Randall Collins and State Breakdown Theory

In order to flesh out what role AI could play in developing social theory, we will now turn to examine one particular social theorist more closely; Randall Collins (see Loyal and Malesevic, 2021: 376-96, for an overview of Collins' work). Our focus on an individual theorist may seem at odds with the scientific method which assesses the validity of hypotheses or theories and not individual thinkers. But as mentioned, it is not Collins as a theorist that we are interested in here, but rather Collins insofar as he put forward hypotheses or theories that can be tested using AI-based models. It can be added that Collins is highly reflective about the scientific nature of the social sciences, including cumulation (Collins and Sanderson 2016: 10-11), how statistics can be regarded as theory (Collins 1984), how simulation can be used for testing theories (Collins 1988: 511-18), and the idea that simulation and other computational approaches should not be regarded as only quantitative but also as qualitative (Collins 1999: Appendix A).

The main illustration that we would like to use here is Collins' prediction of the collapse or breakdown of the Soviet Union (originally published in 1980; see Collins 1999, chapter 2). This prediction was based on models derived from a long line of previous work on state breakdowns such as the French and Russian revolutions (Collins 1999, chapter 1). According to Collins' state breakdown model, there is a non-linear relationship between the size of a state territory, on the one hand, and its success in war, on the other (Collins 1999: p.42). Collins argues that societies with resources and marchland advantage ('marchland' are the boundaries between geopolitical units) are likely to have success in war. At the same time, however, territorial gains lead to increased logistical loads and, eventually to a more central, i.e., exposed, geopolitical position. Hence, expanding empires begin to stall at a certain size, as these points of resistance are aggravated. The details of this hypothesis or prediction about this collapse will not be given here. What is important is that the prediction can be visualized in terms of causal relations (Pearl and McKenzie 2018), spelling out with only a few arrows a 'geopolitical model' of what leads to this collapse (and other prior ones).

*Figure 3. Collins' state breakdown model*

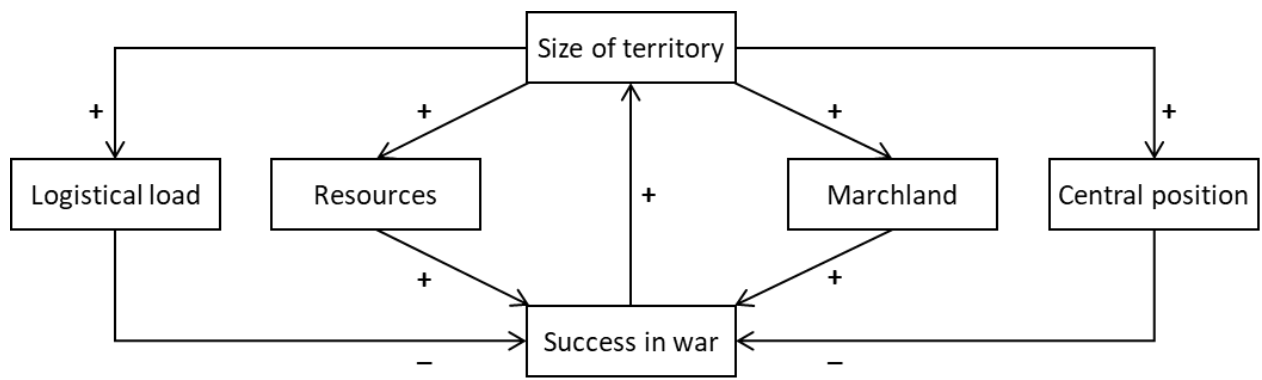

*Figure 3 above is a simplified visual representation of Collins state breakdown model. There are various visualizations of this model with greater complexity, for example in Turner's Macrodynamics (1995: 111). However, Collins' theory of geopolitics can also be formulated in terms of verbal propositions. As summarised by Turner (1991: 251), the collapse of an empire is a positive and additive function of: a) the initiation of war between two empires; b) the over-extension of an empire beyond its logistical capacity; and c) the adoption of its superior technologies by enemy states. While Turner's verbalization has greater complexity and more variables, the essence of the model remains that when resources are stretched too far, the outcome is a crisis of legitimation which, in turn, leads to state breakdown or collapse.*

Collins has discussed the nature of prediction in his model and assessed various alternative predictions and explanations, but he argues that his 'model' fares better. Instead of going into whether Collins outpredicts these rival explanations, our main point here is that Collins' model has been tested by others using computational methods, including sometime collaborators (Hanemann 1987) but also others who build on and extend or improve upon his work (see e.g., Turchin 2019: 16-28) and use his ideas as part of more comprehensive – not just state-breakdown related – testing of theories using computational models. Collins' state breakdown model, as outlined in Figure 3, could thus be inserted in an AI-based model (see Figure 1) as one hypothesis that is ranked against other competing hypotheses depending on their fit with data. Importantly, this process would both be constrained by – and help inform – the evaluation and ranking of social theories about other structures or phenomena at a systems' level (see figure 2, in which geopolitical stability could constitute one of several interdependent 'structures').

It should be remembered that for the Soviet collapse more generally, there were a limited number of explanations. These included, apart from geopolitics, factors such as the lack of openness of knowledge, or the lack of economic growth that provides legitimation for the regime (so-called performance legitimacy, which is also a factor often invoked for the legitimacy of the Chinese party state, see Zhao 2009). Crises in legitimacy have been used to explain state breakdowns drawing on Collins' insights by other scholars, including Goldstone (2020) and Turchin (2019). They, and others, can rely on readily available historical data sets to do so (see e.g., Turchin et al. 2018). They have also made predictions about the future based, among other things, on demographic forecasts. Again, they are able to do this by linking a small number of causal factors, and by setting bounds around the scope of the 'system' to be explained.

At this point, it is worth to briefly pursue a different set of Collins' ideas about AI. In a book chapter (1992) called 'Can sociology create an artificial intelligence?', Collins argued that, indeed, sociology can do this – but it would require a bottom-up approach, i.e., starting with micro-interactions. As mentioned in the introduction, a top-down approach to AI assumes that the rules of thinking need to be formalized in order to be reproduced in AI (for bottom-up versus top-down, see Mitchell 2019). Collins argues that this is the wrong way around: we need to think about how humans learn their social behavior at the micro-level, through social encounters. From the rules that can be extracted from these encounters, and how humans learn those rules and integrate them into their behavior, a model of their thinking can emerge 'from the ground up'. Collins has developed this kind of micro-sociology in a number of papers and books, and we will not spell out the details here. The main point for this paper is that Collins is thereby able to link micro- to macro-; a key issue in theory development in the social sciences.

One example of this micro-macro link is the use of emotional energy as a variable (Collins 2004). On a micro level, emotional energy is increased or decreased through social exchanges. At the same time, emotional energy can also build up and play a critical role at a macro level. One example is violent conflict of large-scale demonstrations that, at critical moments, as during the power struggle between Mikhael Gorbachev and Boris Yeltsin in August 1991, can topple and change regimes (Collins 1999: esp.66). In this case, a series set of events, including emotionally charged demonstrations, led to the removal of Gorbachev and Yeltsin's takeover of power. Here we can see how micro- (demonstrations, with emotional protests which leaders can turn to their advantage using their charisma to delegitimize their rivals and seize power) and macro- (in this case, transfer of power from leader and regime to another) can be linked.

Collins' micro-sociology is testable and measurable (for examples from the context of small group collaboration in science and art, see Parker et al. 2020). Moreover, his theory demonstrates not only how knowledge about social encounters can be built up bottom-up, but also how such knowledge aggregates into theories of social change at the macro-level. The reverse is also true: over the long-term, macro-processes shape the nature of micro-encounters. This has important implications. While some approaches to AI start with 'agents' that are provided with 'rationality' and whose interactions generate social patterns, the examples provided by Collins suggests that such a top-down approach is superfluous. We need not posit features of the model, like rationality, upfront. As we have seen, Collins (2004) starts with emotional energy instead. What Collins' bottom-up approach allows us to do is to theorize (and learn) about how actors behave across the range of social situations, only sometimes producing structural change as with state-breakdowns and revolutions, whereas normally the social order is continually being reproduced, both on the micro- and macro-levels. In other words, it allows us to learn models of routine behaviors, and of the different structural conditions under which these routine behaviors are unsettled, for example, by violent protests (Collins 2009) causing 'situational breakdowns' (Nassauer 2019).

Relating this back to Soviet collapse, the prediction of revolutions or historical turning points of this type is notoriously fraught with difficulty. Micro-social mobilization is indeed sometimes a factor for 'discontinuous' transformations. However, what can be highlighted here is that both routine interactions and the structural transformation of macro- social change provide limits on such aggregated micro-social mobilization that can generate change. Further, on the macro- level, continued legitimation maintains stability unless political instability is caused by events outside the domestic political order such as economic crisis or war (geopolitics). This indicates that micro- and macro- can be linked, as here for politically induced change. But if such linkages are possible to model in the sphere of politics and geopolitics, then it stands to reason that not only the dynamics of other spheres (the economic or cultural spheres, structures 1-3 in figure 2) are model-able, but also the interrelation between these spheres.

In short, there is no need for simulation based on rational agents since we have made the assumption here that only thinking arising from social interactions, including emotions, Collins' bottom-up approach discussed a moment ago, provides the micro-foundations for our approach (although simulation as a complementary approach is not thereby excluded). Nor is there a need for forecasting that sketches out different scenarios or paths: the social world can be modelled with a combination of induction and deduction, with AI-systems testing micro- and macro- linkages over time, and iteratively learning about what outcomes are predicted, and then feeding those predictions, in turn, into new AI-based models. This is arguably what social science does in any case when it is generating new knowledge, though now, in certain areas, with more data and computing power. These AI-systems allow not only analyzing data on a large scale, but also allow rapid iteration of model-building, testing, and in some cases intervening in social processes to influence them and to learn about the outcomes of those influences.

The notion that social science can leverage AI-based models to advance knowledge may be far-fetched because this approach is still at various early stages and has thus far only been applied in limited areas. Moreover, the suggested approach needs further anchoring in ideas about the relation between statistics and quantitative data, as well as about how statistics is theory (Collins 1984) and how 'qualitative' theory complements quantitative inputs and analysis. All these areas have been subject to extensive discussion. But Collins shows that social theory does not need to be based on abstract premises or theories from computing or from the natural sciences – such as evolution or complexity – but can rather be generated from observations of social life and generalizing from these, and from predicting what comes next based on those generalizations. That is, his social thought is premised on sociological or social science knowledge producing findings that can be incorporated into AI-based models such that this learning generates cumulation. Hence, to come back to the point raised at the outset, it is irrelevant to consider Collins as an individual theorist or thinker, except insofar as he produced a version of social science knowledge which, unlike other social science approaches, is free from non- observable assumptions and free from partiality to particular methods or types of data or starting points in computing. His ideas therefore provide the equipment for what could be called machine theorizing.

## 5. The systems we need: strengths and gaps in current AI-systems

The example of Collins, combined with the brief description of the mechanisms underpinning AI-driven social theory earlier in section 3, now allows us to examine if and how advances in AI research can help such an approach to succeed. Looking to the future, adaptive and self-learning AI-systems hold promise for synthesizing knowledge from a variety of sources, reason flexibly and dynamically about the world, apply what is known across a wide range of problems in a systematic way, and transfer what is learned from one context to another (Marcus 2020). However, there is still a long road ahead. Despite recent advances, today's AI-based systems are subject to both technical and conceptual constraints that prevent their full potential from being realized.

From a technical perspective, the sensors used by AI-based systems are still crude in comparison to those of humans. Because keeping track of the state of the world is one of the core requirements of highly potent AI-systems (Russell and Norvig 2015), a more precise and a more diverse set of sensors – or, in the case of social theory, data sources – would enhance the capabilities of AI-based models. Another hurdle preventing the full exploitation of ML techniques are the limits of usable computing power. In a recent article, Di Franco and Santurro (2020) point towards a number of technical limitations associated with the use of artificial neural networks in the social sciences: neural networks require long and resource intensive training before being able to significantly reduce errors in the model fit, and this process requires significant computational resources. However, advances in both sensor technologies and processing power are incremental: both the performance and functionality of digital devices tend to double every second year (Shalf 2020). Hence, our focus here is not on the continuous refinement of known technologies but rather on identifying the *conceptual* gaps in current AI research which – if overcome – would make a qualitative difference to the ability of AI-based models to help advance social theory.

Existing AI-based models have limits in three essential capabilities needed to generate knowledge in ways that are cumulative, holistic, open-ended, and purposeful. These are (1) semanticization, i.e., the ability to develop and operationalize verbal concepts to represent machine-manipulable knowledge; (2) transferability, i.e., the ability to transfer what has been learned in one context to another; and (3) generativity, i.e., the ability to independently create and improve on concepts and models. In the next paragraphs, we will discuss each of these requirements in turn. As we have seen, however, all three can be conceptualized in the approach presented here: what we aim to elucidate are obstacles.

Before doing so, however, two further remarks are appropriate. First, the levels of semanticization, transferability, and generativity displayed by AI-based models are matters of degree. It is thus nonsensical to speak of say, for example, transferability in absolute terms. Even properties of human beings like autonomy and semantic capability exist on a spectrum of behaviour in which they can be compared to the capabilities of other organisms or even machines (Feigenbaum and Feldman 1963). For our purpose of advancing social theory, the AI-system need only have sufficient abilities to transfer relevant knowledge from one domain to another if and when there are mutual dependencies. Sufficient ability to abstract and transfer knowledge may, in this case, mean equal or superior to an adult human being (and likewise with the other two affordances). Second, these properties of AI-based models are not linearly independent. Rather, abilities to semanticize, transfer, and generate knowledge are interlinked and mutually reinforcing. Hence, there is a healthy overlap between different research programs discussed in this section.

Still, and this returns us to current limits, a major shortcoming of current AI-based models is that there is inadequate semanticization. Both natural human language and formal mathematical language are used by social scientists to theorize, and both come with their own sets of strengths and weaknesses (Hanneman 1987). Formal language is more precise and facilitates automated symbol manipulation. However, the ability of AI-based models to represent machine manipulable knowledge in natural human language has direct implications for both system accuracy and robustness: if a system does not master the meaning of the concepts it uses, it will not be able to master the syntax between entities (Gärdenfors 2017). For example, while AI-systems based on deep neural networks can detect correlations between arbitrary bits of information, they lack any internal representation of the external world. This point is worth stressing: neural networks can achieve better performance than linear models if the input data is messy or contains nonlinear relationships (Di Franco and Santurro 2020). However, just because we have data, that does not mean that there are underlying rules that can be learned. Purely correlative systems may therefore get some descriptive theorizing right by drawing on vast databases, but they may not be purposeful since we would not be able to count on them for predictions or to interpret social phenomena that goes beyond the training data (Marcus 2020). Put differently, if our AI-based systems rely solely on curve-fitting and statistical approximation, their inferences will be shallow (Pearl 2019).

Ultimately, useful theorizing is not just about labelling entities; it is about making a coherent interpretation using the best available data. This task requires some degree of reasoning over that data in conjunction with prior knowledge. Hence, if AI-based models are to help advance social theory, they must have the ability to manipulate, and make sense of, symbols at various levels of abstraction. Even in imperfect form, semantic knowledge representation can nevertheless serve as a powerful guide for how to dynamically update models of the external world.

A related shortcoming associated with current AI-based models is the lack of transferability. In computer science, transferability refers to the robustness of a program or a model as it operates in new environments or is fed new data sets. To take the last of these first, the problem of data fusion is currently one of the leading edges of research in big data and computing (Bareinboim and Pearl 2016). However, most current AI-systems are 'narrow' in the sense that they are good at performing a single task (Samoili et al. 2020). While such models may be extremely powerful when applied to the exact environments in which they were trained, they often fail if the environment differs and so they are inherently incapable of adapting to new tasks. For example, while IBM Watson helps clients perform a wide range of tasks from medical diagnostics to customer relations management, the system is not one, but many. For each task, IBM trains a new version of Watson – and the real learning which takes place within each use case remains context specific (Mitchell 2019). The fact that current systems using deep learning often have to begin anew when applied to additional data constitutes a serious limitation for the program of AI-driven social theory as outlined in this paper. It also adds significantly to the time and effort expended on training AI-based models. If AI-systems are unable to generalize knowledge beyond a space of training examples, they cannot be trusted to theorize in open-ended domains. Rather, effective, and feasible transferability builds on drawing inferences from data and from interpolating between situations and entities previously encountered (Marcus 2020). Put differently, transferable AI-based models require mechanisms for learning new knowledge, representing abstract knowledge, and establishing causal links.

Here, we should pause to note that such transferability is also a hallmark of good theorizing: social scientists seek explanations of social phenomena that have some consistency or commonality across domains or local instantiations. Consequently, increased transferability of AI-based models would help advance social theory both by applying known concepts to new contexts but also by testing theories by comparing their validity across multiple dissimilar domains. Efforts in this direction have already been made: Shinn and Joerges (2002), among others, have argued that software acts like passports insofar as it can be applied to many different domains. Here we can, for example, think of visualization tools, or software for social network analysis, but we can also think of how explanatory models and social theory can be applied in different social science domains.

Finally, for AI-systems to advance social theory more effectively, it would be useful if they could also help generate concepts and theories. Generativity can be understood as the ability to use known data and transform it so as to create something new and potentially surprising (Compton and Mateas 2017). For example, while simple pattern recognition may help classify input images into pre-defined output categories, generative AI-based models would be able to conceptualize new output categories based on a sufficiently large set of input images. According to Smith (2019), this is a real possibility since ontology emerges in the context of registration. Assuming that there will always be a surplus to that which is being registered, AI-systems could help us understand what the world is like beyond the concepts hitherto used by social scientists by registering the social world in new and different ways.

However, this does not necessarily imply that the AI-system itself needs to be 'creative' in the human sense. Rather, generativity is a property of the process whereby AI-based models are expanded, augmented, or transformed so as to best fit new data with previous knowledge. Admittedly, this kind of generativity presupposes that AI-based models already possess certain levels of semanticization and transferability. However, while these criteria are necessary, they are not sufficient: generativity also requires that rules and exceptions must be allowed to coexist (Marcus 2020). As a case in point, minimizing errors in ML models based on past data is well known to cause overfitting and result in poor predictive performance (Dietterich 2017). As Radford and Joseph (2020) put it, feature engineering risks boosting model performance (narrowly defined) without helping us to distinguish genuine patterns in the data. Thus, just as no amount of upfront programming can substitute for learning, the capacity to 'learn' is not enough for a system to generate new hypotheses. To generate new concepts and theories, AI-based models would need to build on initial frameworks for domains like time, space, and causality in order to guide – and speed up – subsequent learning.

Stinchcombe (1968) noted that if a sociology student has difficulty thinking of at least three sensible explanations for a correlation under investigation, he or she should probably look for another profession. While this may seem a high bar, no lower standards should be expected of AI-systems if they are to help advance social theory. We have already seen examples where ranking explanations (where there are typically a few top contending explanations) can be applied to advance theories of social change. As mentioned, the ability to 'come up with explanations' requires generativity, and to ensure that they are 'sensible' (or indeed explanations at all) requires semanticization and transferability. Hence, what may be needed is to develop AI-systems that include features of various AI technologies, including symbolic expert systems and artificial neural networks (Marcus 2020). The idea of so called 'hybrid systems' is not new, and teams of computer scientists (amongst others at Deep Mind in London) are currently working in that direction. It is thus a real possibility that architectures that combine large-scale learning abilities with symbolic knowledge representation can enable AI-based models to acquire, represent, and manipulate abstract knowledge, and use that knowledge in the service of building and updating theories of complex social phenomena.

## 6. Limitations of AI-driven Social Theory

The approach suggested in this paper has been in keeping with the four principles of good AI design outlined in section 2. To reiterate, these stipulate that a good AI-model of social change should be cumulative (i.e., allow only scientific criteria to guide knowledge creation), holistic (i.e., encompass both macro- and micro-dynamics in relevant disciplines), open-ended (i.e., allow for the inclusion of ever-more and improved data sources, and ways to incorporate them), and purposeful (insofar as ideas are provided about how to improve upon the workings of existing technologies and existing social theory). Nevertheless, the programme sketched in this paper is also subject to a range of limitations. Some of these have already been mentioned; namely, that both the data that can be used by AI-systems, and also how the data relates to the social world, are limited, and skewed to specific populations or phenomena. We now discuss some further limitations and potential objections – as well as how ongoing developments may alter the conditions for AI-driven social theory going forward.

A key issue in using AI techniques is data availability. One reason why AI-driven social theory has recently gained steam is because much more data has become available. But these recent data are also limited; they tell us only about social behavior that is captured digitally. These data largely pertain to digital media, and this characteristic prompts us to consider the limits of what media reveal about social life: Media constitutes a limited but growing part of social activity even if we include among media, for example, CCTV footage and the like. That this is a limitation can be easily grasped by the fact that such data goes back only a quarter of a century or so. However, the digitization of knowledge concerning social phenomena through books and other sources goes much further back, and so too does data about a range of social activities, such as economic activity. In certain cases, such historical data is even being made manipulable by means of providing ontologies (Peregrine et al. 2018). This availability of ontologies and digitization provides new opportunities for AI-driven theory development. As Eagle and Greene (2014) have shown, social behavior can also be digitally captured all the way from the micro- to the macro-levels; or, in their account, from lifelogging the details of an individual's everyday behavior to mining the behavior of the world's population.

A further obstacle at the intersection between computer science and social science is that the availability of data for different terrains or areas of social life is highly uneven – as are the social theories pertaining to them. We have limited ourselves to the social sciences concerned with social interaction or structures. These include political science, sociology, and anthropology. Psychology was excluded insofar as it is concerned with the cognitive development of individuals, and economics insofar as it is based on isolated individual interests. In other words, the three disciplines included here are concerned with social relations or structures rather than with individual minds or preferences. It can be mentioned parenthetically that AI, historically, has most closely been associated with psychology or cognitive science. This is unsurprising since one aim of AI has been to reproduce thinking. However, we have touched on this only briefly with reference to the sociology (not psychology) of thinking earlier.

Social science will continue to make the most rapid advances based on the uses of information and communication (media) devices because data are most readily available and manipulable from these sources (more on this in section 7 below). But contra the cheerleaders of computational social science, it needs to be remembered that media uses play a limited role in social change. On the other hand, media do play a large role in social change and in the social sciences in the sense that media set the agenda for politics and sometimes for other social changes. Another shift in data sources that is bound to take place is that a greater proportion of digital data will come from the Global South, where mobile phones play an increasing role and the data from these devices is often available. However, this availability also points to further limits: while digital media data are more abundant, they do not necessarily provide representative samples. China's population, for example, is in many respects missing as a data source because its data mainly come from platforms that are separate from those in other parts of the world and, even more critically, often not available to researchers (though this is of course also a major issue outside of China). Other missing data include the 'less media-ted' parts of the world's population or of the social world, though this is a problem that can partly be addressed by 'weighting' this part of this population against others for which data are available.

Further, disciplinary divides and divides in technoscientific development will hamper advances in AI-driven social theory, while cross-disciplinary collaboration will accelerate them. There are many promising research projects in the social sciences but no coherent programme, and the same applies to the development of databases, ML algorithms and other AI technologies. Another problem that will hamper advancements is potential resistance among academics and AI researchers. Such resistance may, in the case of social scientists, take the shape of skepticism against technological determinism or social engineering (which may also spill over to a wider public), and in the case of AI researchers, a hesitation toward, or lack of incentives for, engaging with the social sciences. AI developers may also face backlashes from those who wish to curb implementation of the technology. However, while these forms of resistance may slow the utilization of AI-driven social theory or channel it into certain directions, they will not ultimately impede the advancement of social science knowledge.

This resistance relates partly to the question whether AI-driven social science knowledge can move from knowledge to practice; for example, to identify where political actors can most effectively intervene to address social problems. The issue is that there are many such social problems. Turner (2019) distinguishes between applied sociology, or what he calls social engineering, on the one hand, and activism or ideologically-driven sociology, on the other. Scientific social science, in his view, should not become the handmaiden of political activism: this subjugates knowledge advancement to political goals and so can lead it away from cumulation and being scientific. However, practice-driven social engineering is already taking place partly due to big data: Karpf (2016), for example, has shown how social movements are driven by 'social listening' to what the members of social movement organizations want. Such efforts to yoke social listening to political causes can lead social science away from going beyond the most advanced state-of-the-art knowledge or the most reliable synthesis of existing knowledge. That is because they seek to focus on what is politically most relevant to particular interest groups which, in turn, can give the public the impression that social science is not trustworthy but rather politically driven.

One objection to AI-driven social theory is indeterminacy: if the AI-driven social theory is encompassing, then what room is there for incorporating new unforeseen elements? However, if an element is truly unforeseeable, the only answer would be to wait until this eventuality arises. A related but separate issue is 'agency': with an all-encompassing deterministic theory, what scope is there for freedom or for policy that is not coupled to the knowledge that has been gained? This is an age-old problem in the social sciences. But granted that there is freedom in how to apply social science knowledge, there is also 'reflexivity' in relation to how new knowledge is coupled with phenomena and 'awareness' about the link between knowledge and its translation into practice. What is interesting is that the same problem has been raised in relation to AI, except that it relates to the alleged 'determinism' of applying knowledge to the application of technology (the famous driverless car and its choices about which pedestrians to run over, the 'trolley problem'). But here too there are solutions that rest on coupling knowledge to ethical behavior. These problems were anticipated by Kant: deterministic knowledge need not be in conflict with the free will that is the basis of ethics (Gellner 1974: 184-188).

Apart from the complexity and reflexivity of social science, it could also be objected that science is too complex to allow for the kind of reduction that the AI-driven approach to social theory put forward here would imply. However, science is not somehow ineluctably complex: Hacking (2012), for example, has argued that there are six fundamental styles of scientific knowledge. Without going into these here, all six styles of scientific knowledge can apply to social science, and, for each style, the corresponding data sources are available (Meyer and Schroeder 2015). If, furthermore, and again following Hacking, science consists of representing and intervening (Hacking 1983), then the interlocking adventure between them (to continue with Hacking's terminology), whereby technoscience represents phenomena and is able to manipulate them, can also be sketched for AI-driven social theory. In the case of AI, this interlocking relates to the iterative testing of representations of the social world and how they are coupled to it, and so allow the AI-system to learn.

Another objection to the approach proposed here could be: why should social scientists bother with AI, given that data-driven AI efforts concerning far more circumscribed parts of the social world still have teething problems? Yet the question only needs to put in this way in order to recognize a number of interrelated and obvious replies: first, why should AI-driven social science be any more or less tractable in tackling social theory than in more delimited domains such as predicting crime or violence? Second, if AI is simply what thinking humans cannot easily do, then social theory is perhaps no different from other areas where machines are able to do better than humans. Third, if it is possible to visualize the components of a social theory and identify a logic of explanation (Pearl and McKenzie 2018), then devising a machine-manipulable approach will be a useful scientific start. Fourth, macro- theory, which could be regarded as the most complex of social theory of the social world, has made starts in the direction of AI-based models, and since the macro- micro- link has been addressed by a number of social scientists and sketched here (without pre-judging the success of this link), there is no reason why the same link should not be incorporated in an AI-driven social theory.

## 7. Outlook for AI-driven Social Theory

Despite these limitations and objections, is it nevertheless possible to harness knowledge concerning the most pressing social needs in an objective way? One blueprint from the natural sciences are the attempts by Rockstroem and colleagues (Lade et al. 2020) to identify 'planetary boundaries'. This model works, as it were, backwards from the planetary boundaries to the most urgent tasks to stay within these boundaries in terms of resource conservation. Such an effort would entail prioritizing which of these efforts are likely to yield the maximum contribution in doing so. Perhaps something similar could be done with social science knowledge? If so, social theory could be directed at the most pressing issues, thereby reducing uncertainty. AI-driven social theory could then take into account the particular time-, space-, and resource horizons of social actors (including organizations and macro-actors) who (or which) aim to intervene in social life, and scientifically gauge the potential effect they can have in view of the predictability of what can happen next.

Which actors should be informed by what type of knowledge can itself be made subject to a learning process, and thus made more effective via a pipeline of knowledge inputs and outputs. However, again, such efforts must be scientific and steer away from activism and instead be problem-oriented. In any event, support for particular actors or social engineering is secondary to the goal of the observer of society; the social theorist or theories whose only goal is knowledge advance in relation to particular social phenomena - as opposed to the problem-oriented actor.

This is a good place to distinguish between prediction, as discussed in this paper, and global forecasting (see, for example, https://pardee.du.edu/). Forecasting is not cumulative unless it can be integrated into more useful and reliable social science knowledge. That is because forecasting has various 'scenarios' for the which the parameters vary. The same holds true for simulations, which are based on the constraints that social actors have to behave in certain ways, which can also be seen as forecasts. These varying parameters also introduce trade-offs in terms of how much the analyses explain. Obviously, the same does not apply to the example of prediction, such as Collins' prediction of Soviet collapse presented here (though it would apply if his prediction was related to future state breakdowns which include different 'scenarios'). Does AI-driven social theory always aim at prediction? Not necessarily, since it may also seek non-repeatable patterns in the past. But it does so by means of techniques (e.g. visualizations and formalizations) that can be applied to new circumstances or domains, in breadth or depth or in the future, and make claims about these. In that sense, it still seeks to predict (or 'retrodict').

One topic that could provide another example of the approach outlined here and of addressing pressing social needs is inequality. This is a topic where much social science knowledge has (ac)cumulated, and yet existing theory can be improved. One way to examine inequality would be with the practical aim of promoting solving social problems, though inequality could equally be examined without such an aim since the topic also has intrinsic theoretical merit in understanding and explaining social change. Knowledge about inequality has often been global and long-term. Naturally, digital media data has been of limited use here since much work concerns the past. However, measurement of inequality has been extensive and is continually being refined (Milanovic 2016) and AI-systems could – if designed in line with the principles stipulated in section 2, and structured according to the model outlined in section 3 – be used to further improve existing theories based on the available data.

More specifically, AI-driven social theory could help modelling the relationship between economic growth and income or wealth inequality, as well as predicting the relationship between them – given certain assumptions about changes in the rates of economic growth (Maddison 2007). There is therefore no reason in principle – such as, for example, the complexity of the social world or the like – why this relationship should not be model-able: Piketty (2014: 52) has formulated the 'first fundamental law of capitalism', thereby reducing inequality to a formula. The data are abundant, and their limitations are well-known. There are other mathematical models, including models which engage with the work of Piketty and with general social theories such as Turner's (of whom more in a moment) and which address the more multi-dimensional aspects of inequality, and also tie them to model-able ideas of justice (for example, Jasso 2021). Such efforts could be separate from Collins' state breakdown model since economic power operates via different mechanisms – though they could be combined in a single AI-based model if state breakdown (or failure) leads to decreasing inequality or 'leveling' (see, for example, Scheidel 2018). Turner, for example, has provided a general theory of society which incorporates Collins' geopolitical model (1995: 108-13) and also links it to the workings of inequality (or what Turner calls 'distribution dynamics'). Turner furthermore links the macro-dynamics of this general theory to micro- and meso-processes in other work. In short, the theoretical elements for aggregating AI-based models as per our figure 2 above are readily available, though implementing this approach would take considerable work.

A further prediction can be made based on Collins' ideas as presented in this paper. This prediction states that social science will advance most rapidly and cumulatively in the realm of (mainly digitally) mediated interaction, because that is where the vast bulk of computationally manipulable data come from (though, as Brent [2017] reminds us, there are plenty of databases for traditional or legacy media, and for non-media related social process such as historical events and many more: again, Brent is only one source for the many databases available for analyses in AI theory building, which cannot be detailed here). It is also possible to spell out the obstacles to progress in this area: poor integration of disciplines, and low mutual dependence and task uncertainty (Whitley 2000; Schroeder 2007). But rising above these limits, from the side of social science, it is clear that certain dynamics are particularly model-able. This, combined with the fact that a seamless (if under-theorized) integration of micro- and macro- allows for setting the macro- constraints for micro-interaction (as in geopolitics, as we have seen) as well as for aggregating micro- interactions to build up to macro-trends, improves the prospects to advance social science. Efforts at applying such dynamics to research on digital media are already emerging. For example, to stay with those applying the ideas of Collins, DiMaggio et al. 2018 have done this, in their case at the meso-level.

Yet another way to refine – and learn from – the state breakdown model of Collins and others would be test it for other powers such as China and the US. Here it should be stressed that 'breakdown' is not necessarily the same as 'breakup', as with the Soviet state: breakdown in China or the US could occur with a failure of outside-in (geopolitical) state legitimacy, leading to a radical change of politics (from authoritarian to democratic, or the other way around). Again, a data-driven attempt could be made to find an AI-based model of the translation mechanism(s) between geopolitics and economic growth or inequality – and vice versa.

As we have seen, it has been proposed that sociology can help technoscientific advance in developing AI by enabling a computer to learn the rules of sociability at the micro-level and thus engaging with the social world. This 'bottom-up' approach could lead to machines functioning as social human beings. If such machine-approaches can be developed, and more powerful social science knowledge developed with the help of AI-systems, then ultimately the two research programmes may connect: more close-to-social human behavior at the micro- and macro-levels can lead to more powerful thinking machines, and more powerful thinking or knowledgeable machines can model and manipulate the world more readily. This potential meeting point can be left open; nothing hangs on it for the argument here.

In social theory, if the point of departure is the top-most macro-level (including the relevant phenomena at the meso- and micro-level that contribute to macro-level change), then there are currently often a limited number of alternatives: there is modernization theory with its implicit evolutionism;

Marxism and world-systems theory; and geopolitical theory. To this could now be added Anthropocene theory, with its environmental determinism. If conflict and cohesion are at the root of these theories, and unit size (nation-states, or a global system) is given, and we leave aside for the moment 'culture' or 'cultural determinism' which depends on media as a vehicle (or if media are regarded not as a separate system but a subsystem), then the range of options in social theory, again, is quite restricted. That could be a strength since the top-down model presented in figure 2 (see section 3) could revolve around testing these few alternatives, eliminating those that do not fit, and moving on to other theories. In this way, thinking machines may help social theory to conquer territories and move on to new ones.